\documentclass[
  reprint,
  superscriptaddress,
  nofootinbib,
  amsmath,amssymb,
  aps,
  prd
]{revtex4-2}
\usepackage{amsmath}
\usepackage{graphicx}
\usepackage{comment}
\usepackage{xcolor}
\usepackage{orcidlink}
\usepackage[normalem]{ulem}
\usepackage{hyperref}

\begin{document}

\title{How deep can a cosmic void be? \\Voids-informed theoretical bounds in Galileon gravity}

\author{Tommaso Moretti\,\orcidlink{0009-0006-4815-4764}}
\email{tommaso.moretti.pv@gmail.com}
\affiliation{Dipartimento di Fisica “E. Pancini”, Università degli Studi di Napoli “Federico II”,\\Compl. Univ. di Monte S. Angelo, Edificio G, Via Cinthia, I-80126, Napoli, Italy}
\affiliation{INFN Sezione di Napoli, Università degli Studi di Napoli “Federico II”,\\Compl. Univ. di Monte S. Angelo, Edificio G, Via Cinthia, I-80126, Napoli, Italy}

\author{Noemi Frusciante\, \orcidlink{0000-0002-7375-1230}}%
\affiliation{Dipartimento di Fisica “E. Pancini”, Università degli Studi di Napoli “Federico II”,\\Compl. Univ. di Monte S. Angelo, Edificio G, Via Cinthia, I-80126, Napoli, Italy}
\affiliation{INFN Sezione di Napoli, Università degli Studi di Napoli “Federico II”,\\Compl. Univ. di Monte S. Angelo, Edificio G, Via Cinthia, I-80126, Napoli, Italy}

\author{Giovanni Verza\,\orcidlink{0000-0002-1886-8348}}
\affiliation{ICTP, International Centre for Theoretical Physics, Strada Costiera 11, 34151, Trieste, Italy}
\affiliation{
Center for Computational Astrophysics, Flatiron Institute, 162 5th Avenue, 10010, New York, NY, USA
}

\author{Francesco Pace\, \orcidlink{0000-0001-8039-0480}}
\affiliation{Dipartimento di Fisica, Università degli Studi di Torino, Via P. Giuria 1, I-10125 Torino, Italy}
\affiliation{INFN-Sezione di Torino, Via P. Giuria 1, I-10125 Torino, Italy}
\affiliation{INAF-Istituto Nazionale di Astrofisica, Osservatorio Astrofisico di Torino, strada Osservatorio 20, 10025, Pino Torinese, Italy}

\begin{abstract}
We establish a void-based consistency test for Galileon scalar-tensor theories. We show that the previously reported unphysical breakdown of the predicted Newtonian force in certain Galileon models is controlled by a single condition linking non-linear void dynamics to the cosmic expansion history. This connection yields a redshift-dependent upper bound on the allowed depth of voids and promotes this requirement to a new viability condition, complementary to standard stability criteria. As an example, we apply this void-based criterion to a linear parameterization in the scale factor constrained by theoretical and observational bounds; we find that $\sim 60\%$ of the parameter space is excluded, with most problematic models failing by $z\lesssim 10$. These results position cosmic voids as sharp, complementary and theory-informed filters for viable modified gravity, enabling more informed priors and parameter-space choices in future cosmological inference. 
\end{abstract}

\maketitle

\section{Introduction}
The $\Lambda$CDM model remains the standard framework for interpreting cosmological observations, but the extreme fine-tuning of the cosmological constant $\Lambda$~\cite{Bull:2015stt} and emerging tensions between low- and high-redshift measurements~\cite{CosmoVerseNetwork:2025alb} motivate extensions beyond General Relativity (GR). Modified gravity (MG) scenarios in which cosmic acceleration is driven by new dynamical degrees of freedom
offer a particularly well-defined avenue, as they generically predict correlated signatures in both the background expansion and the growth of structure~\cite{Joyce:2014kja,Lue:2004rj,Copeland:2006wr,Silvestri:2009hh,Nojiri:2010wj,Tsujikawa:2010zza,Capozziello:2011et,Clifton:2011jh,Bamba:2012cp,Koyama:2015vza,Avelino:2016lpj,Joyce:2016vqv,Nojiri:2017ncd,Ferreira:2019xrr,Kobayashi:2019hrl}. Among these, scalar-tensor theories of the Galileon type~\cite{Horndeski:1974wa,Deffayet:2009mn} characterized by non-linear derivative self-interactions, have attracted sustained interest because they can generate cosmic acceleration without introducing a bare cosmological constant~\cite{Silva:2009km,Deffayet:2010qz,Kobayashi:2010cm} and because they admit a screening mechanism~\cite{Vainshtein:1972sx} that reconciles large-scale modifications with stringent local tests of gravity.

However, the enlarged parameter space introduced by an extra degree of freedom is strongly constrained by viability. The stability of the scalar and tensor sectors~\cite{Frusciante:2016xoj,DeFelice:2016ucp,Frusciante:2018vht} imposes ghost, gradient, and  tachyonic conditions on the parameter space, which must be enforced prior to cosmological inference~\cite{Raveri:2014cka,Salvatelli:2016mgy}. Moreover, the phenomenology of such models is tightly restricted by theoretical consistency and by multimessenger and astrophysical observations: (i) the $\mu$--$\Sigma$ conjecture~\cite{Pogosian:2016pwr} links modifications of structure growth and light deflection, preventing viable Galileon models from altering clustering and lensing independently; (ii) the near-simultaneous detection of GW170817 and GRB170817A requires the gravitational wave (GW) speed to be extremely close to that of light~\cite{LIGOScientific:2017zic}; (iii) dark energy (DE) perturbations may become unstable in the presence of a GW of sufficiently large amplitude, then constraining braiding effects~\cite{Creminelli:2019kjy}; and (iv) strong-field effects further constrain departures, as the piercing effect~\cite{BeltranJimenez:2015sgd} weakens Vainshtein screening near compact objects, making Galileon theories particularly sensitive to tests of anomalous GW propagation. Together, these criteria shape the region of observationally viable MG theories. 

Cosmic voids emerge as sensitive probes of MG~\cite{Perico:2019obq,Clampitt:2013tyg,Cai:2014fma,Davies:2019yif,Barreira:2015vra,Voivodic:2016kog,Paillas:2018wxs,Maggiore:2025mbp,Moretti:2025gbp,Takadera:2025ehm,Moretti:2026dfz}. As large, underdense environments where screening mechanisms are inefficient, voids amplify deviations from GR. Thus, voids complement traditional large-scale structure probes.
In some Galileon models, the predicted Newtonian force inside the voids can become unphysical~\cite{Barreira:2013eea,Winther:2015pta,Baker:2018mnu,Li:2013tda}. This pathology can lead to ill-defined particle dynamics and unreliable void observables~\cite{Baker:2018mnu}.

In this paper, we show that this pathology can be avoided by imposing a simple condition defined only by background functions, which yields a redshift-dependent upper bound on void depth and introduces an additional viability requirement during structure formation, complementary to the criteria presented before. This void-based constraint reshapes the allowed parameter space, promotes voids as sharp, theory-informed filters of viable MG, and enables more informed parameter space choices in future cosmological inference.

The paper is organized as follows. In Section~\ref{Sec:Modified_Newtonian_force}, we introduce the modifications to the Newtonian potential in Galileon scalar-tensor theories, derive the corresponding non-linear effective coupling inside spherical voids, and discuss the pathological regime of the fifth force. In Section~\ref{Sec:Voids-informed_criterion}, we present our void-informed consistency criterion to identify the viable region of parameter space and compute how deep a cosmic void can be. In Section~\ref{Sec:results}, we apply these two diagnostics to a case study to infer the excluded and viable regions. Finally, in Section~\ref{Sec:conclusions}, we summarize our findings and discuss their implications.

\section{Modified Newtonian force}\label{Sec:Modified_Newtonian_force}
At the non-linear level, the Newtonian gravitational potential, $\Psi (t, {\bf x})$, is related to the non-linear density contrast, $\delta(t, {\bf x})$, through the Poisson equation.
MG models introduce an additional fifth force beyond Newtonian gravity, typically mediated by the scalar field interaction, that provides an extra contribution to the potential. Then, modifications to the gravitational strength can be encapsulated in an effective coupling $\mu_{\rm NL}(t,\mathbf{x})$, leading to a modified Poisson equation
\begin{align}
    \frac{\nabla^2}{a^2}\Psi = 4\pi G\,\mu_{\rm NL}\,\bar{\rho}_{\rm m}\,\delta\,,
\end{align}
where $a(t)$ is the scale factor of the universe, $G$ is the Newtonian constant, and $\bar{\rho}_{\rm m}$ is the background density of the matter components, i.e. baryons and cold dark matter (CDM). We note that the function $\mu_{\rm NL}$ has to encode non-linear effects associated with screening.
This approach allows the function $\mu_{\rm NL}$ to be parameterized in a model-independent way by choosing its functional form, ensuring that a screening mechanism is included. However, in this case, the connection to a specific underlying theory of gravity is lost. In contrast, assuming a specific gravitational theory, it is possible to identify the corresponding functional form for $\mu_{\rm NL}$.

Here, we work with Galileon scalar-tensor theories, using the time-dependent $\alpha$-basis~\cite{Bellini:2014fua}: $\{\alpha_{\rm K},\alpha_{\rm B},\alpha_{\rm T}, \alpha_{\rm M}\}$, characterizing, respectively, the scalar kinetic energy, kinetic braiding, tensor-speed excess, and the evolution of the effective Planck mass, $\alpha_{\rm M}\equiv {\rm d}\ln M^2/{\rm d}\ln a$, with $M$ being the effective Planck mass. Guided by the stringent multimessenger bounds on $|\alpha_{\rm T}|\lesssim10^{-15}$, we set $\alpha_{\rm T} = 0$ at all redshifts~\cite{Bettoni:2016mij,Creminelli:2017sry,Ezquiaga:2017ekz,Baker:2017hug,Sakstein:2017xjx}. In general, $\alpha_{\rm K}$ can be fixed to a constant value, given that its contribution to the observables is below the cosmic variance~\cite{Frusciante:2018jzw} and cannot be constrained by data~\cite{Bellini:2015xja}. In our analysis $\alpha_{\rm K}$ does not appear in any expression so we do not need to specify it. Physically relevant deviations from GR are therefore fully specified by $\alpha_{\rm B}$ and $\alpha_{\rm M}$.  

For these theories, and under the quasi-static (QS) approximation, valid on sub-horizon scales where the time derivatives of the perturbations are subdominant compared to the spatial gradients, we use the equations derived in~\cite{Kimura:2011dc}. Then, we assume a homogeneous, isotropic, and flat background, defined by the Friedmann-Lema{\^i}tre-Robertson-Walker background metric and assume spherical symmetry for the perturbations. In this case, the non-linear solution for the gravitational potentials and the Galileon field depends on the matter source only through the enclosed density contrast $\Delta$.
We adopt the inverse top-hat density profile to model voids under-density, only as a simple illustrative choice~\footnote{The reasoning here and the explicit expressions do not require a specific top-hat profile, but they can be more generally written in terms of the average enclosed density contrast.}, and we compute the non-linear modification to the effective gravitational coupling, which takes the form
\begin{align}
    \mu_{\rm NL}(a,R) &\,=\, \frac{M_{\rm pl}^2}{M^{2}}\Bigg\{1
    + 2\left(\frac{M^2}{M_{\rm pl}^{2}}\mu_{\rm L} - 1\right)
    \left(\frac{R}{R_{\rm V}}\right)^{3}
    \nonumber\\
    &\quad \times\bigg[\sqrt{1+\left(\frac{R_{\rm V}}{R}\right)^{3}}- 1\bigg]\Bigg\}\,,
    \label{eq:non_linear_gravitational_potential}
\end{align}
where $M_{\rm pl}$ is the Planck mass and
\begin{align}
    \mu_{\rm L}(a) \,=\, \frac{M_{\rm pl}^2}{M^{2}}\left[1 + \frac{2\left(\frac{1}{2}\alpha_{\rm B} + \alpha_{\rm M}\right)^{2}}{\alpha\,c_{\rm s}^{2}}\right]\,,
    \label{eq:linear_potential}
\end{align}
is the linear modification of the gravitational strength. We also define
\begin{align}
    \alpha\,=\,\alpha_{\rm K}+\frac{3}{2}\alpha_{\rm B}^2\,,
\end{align}
and $c_{\rm s}^2$ is the scalar sound speed~\cite{Bellini:2014fua}:
\begin{align}
    \alpha\, c_{\rm s}^{2}\,=\, -(2 - \alpha_{\rm B})\left[\frac{H'}{H} - \frac{\alpha_{\rm B}}{2} - \alpha_{\rm M} \right]
    + \alpha'_{\rm B} - \frac{\bar{\rho}_{\rm m}}{M^2}\,,
\end{align}
where $'\equiv\partial/\partial\ln{a}$ and $H\equiv ({\rm d}a/{\rm d}t)/a$ is the Hubble function. For theoretical stability requirements, ghost and gradient instabilities are avoided by setting  $\alpha\,>\,0$ and $c_{\rm s}^2\,>\,0$, respectively.
Finally, $R$ is the physical radius of the void shell, and $R_{\rm V}$ is the Vainshtein radius, the characteristic screening scale below which the fifth force is suppressed and  GR is restored on small scales while allowing modifications on large, low-density scales. Their ratio is
\begin{align}
    \left(\frac{R_{\rm V}}{R}\right)^{3} = \frac{
    4(\alpha_{\rm B} + \alpha_{\rm M})(2\alpha_{\rm M} + \alpha_{\rm B})\,\Omega_{\rm m}}{\frac{M^{2}}{M_{\rm pl}^2}\left(\alpha\,c_{\rm s}^{2}\right)^{2}}
    \,\delta\equiv f_{\rm MG} (a)\,\delta\,,
    \label{eq:RV_over_R}
\end{align}
with $\Omega_{\rm m}(a)= \bar{\rho}_{\rm m}/3 M_{\rm pl}^2H^2$. Here, for diagnostics, we define the function $f_{\rm MG}(a)$ which is a pure background function. More generally, in Eq.~\eqref{eq:RV_over_R} the quantity $\delta$ should be understood as the enclosed density contrast $\Delta$, i.e. the spherical average within radius $R$, so that the discussion can be formulated without relying on a strict inverse top-hat profile.

Eq.~\eqref{eq:non_linear_gravitational_potential} clearly shows that $\mu_{\rm NL}$ remains real only if the square-root argument is non-negative. In voids, where the enclosed matter contrast is negative, the ratio $(R_{\rm V}/R)^{3}$ can be negative, and whenever it drops below $-1$, the square root becomes imaginary and hence $\mu_{\rm NL}$, leading to an unphysical gravitational strength. This argument is controlled by $\Delta$ and therefore does not depend on whether the void is described by an exact inverse top-hat profile or by a more general spherically symmetric profile.  
This is the basic origin of the pathological behavior that we exploit as a consistency criterion. 
In spherical collapse analysis, used to model halos, a similar problem arises but only for Galileon scenarios with $\alpha_{\rm T}\neq0$~\cite{Barreira:2013xea,Kimura:2011dc}, where the same square-root structure appears in the non-linear force. For voids, this pathology has already been reported in several studies, all focusing on the self-accelerating branch of Galileon gravity in which the scalar field is responsible for the late-time accelerated expansion~\cite{Kimura:2011dc,Baker:2018mnu,Barreira:2013eea,Winther:2015pta}. When this issue appears in $N$-body simulations, an ad hoc but practical prescription is typically applied to keep the force real~\cite{Barreira:2013eea}. To clarify the nature of this instability, a key theoretical assumption was re-examined, namely the QS approximation used to derive the fifth force (and $\mu_{\rm NL}$). Solving the full time-dependent field equations for both over- and under-densities shows that the QS solution is an attractor and that the approach to a negative square-root argument triggers runaway behavior, pointing to a genuine instability of the model rather than a breakdown of the QS ansatz~\cite{Winther:2015pta}.

\section{Voids-informed criterion}\label{Sec:Voids-informed_criterion}

In this work, we introduce a simple, physically motivated criterion to identify the region of parameter space in which the MG force remains well defined in the QS, spherically symmetric void configurations considered here.

Using this criterion, we delineate the viable parameter space and determine the corresponding range of predicted void depths. This provides a robust consistency check that complements existing theoretical and observational constraints and clarifies which model realizations yield physically meaningful void phenomenology.

We require the square-root argument in Eq.~\eqref{eq:non_linear_gravitational_potential} to be non-negative for all void configurations ($\delta \ge -1$):
\begin{align}
    1+f_{\rm MG}\,\delta\geq 0\,,
    \label{eq:viable_sqrt}
\end{align}
thereby ensuring a real non-linear coupling, $\mu_{\rm NL}$, and a well-defined fifth force. 
We consider that sufficiently underdense regions can reach $\delta\simeq -1$, so any model for which $f_{\rm MG}$ exceeds unity at any redshift during the structure-formation era necessarily admits  configurations with $f_{\rm MG}\,\delta<-1$, for which $\mu_{\rm NL}$ becomes imaginary and the theory loses viability. This argument translates into the following criterion
\begin{align}
    \max_{0 \leq z \leq z_{\rm in}} f_{\rm MG}(z) > 1
    \quad \Rightarrow \quad \text{model excluded}\,,
    \label{eq:max_criterion}
\end{align}
where $z_{\rm in}$ is the initial redshift. Here, we choose $z_{\rm in}=100$ to encompass the epoch relevant for structure formation and match the regimes probed by $N$-body simulations. The criterion in Eq.~\eqref{eq:max_criterion}, defined only by a time-dependent background function, can be imposed before any prior knowledge of the non-linear dynamics to identify the viable parameter space of a given model.
The exclusion criterion in Eq.~\eqref{eq:max_criterion} is not only mathematically well defined, but also physically motivated. From a theoretical perspective, its meaning is clear: once $\max f_{\rm MG}>1$, there exist underdense configurations for which the pathological branch is inevitably encountered. We stress that, when applying the criterion to simulations or observations, this is not tied to any specific operational definition of a void, but depends only on the enclosed density contrast $\Delta$. The variance of the smoothed matter density field is a monotonic decreasing function of the smoothing radius. As a consequence, on large scale, e.g.\ $R\sim 100\,h^{-1}{\rm Mpc}$, the smoothed density contrast value $\Delta$ approaches 0. However, smoothing the density field around the center of a cosmic voids at small scales, e.g. $R\sim 1\,h^{-1}{\rm Mpc}$, can lead to a density contrast value approaching $\Delta \sim -1$~\cite{Bayer:2021iyb,Verza:2024rbm}. For this reason, excluding models as soon as $\max f_{\rm MG}>1$ is not overly conservative: even voids that are not particularly deep on large scales can contain physically plausible inner regions where the pathological branch is activated and the fifth force becomes ill defined (see, e.g.,~\cite{Barreira:2013eea}).

Despite the fact that our derivation is performed under spherical symmetry, this criterion remains relevant more generally. First, the inner cores of voids, which correspond to their deepest regions, are expected to be closer to spherical, while the outer regions are more strongly shaped by the surrounding environment. Second, the argument is formulated in terms of enclosed averages, so departures from spherical symmetry are not expected to regularize a pathological root, although we do not demonstrate this here.
 
Additionally, $f_{\rm MG}$ determines for each $z$ the deepest under-density that can be realized without entering the imaginary branch. The requirement in Eq.~\eqref{eq:viable_sqrt}, combined with the physical bound $\delta \geq -1$, implies that the minimum allowed density contrast at a given redshift is
\begin{align}
    \delta_{\rm min}(z) \,=\, \max\!\left(-1\,,-\frac{1}{f_{\rm MG}(z)}\right)\,.
    \label{eq:minimum_delta_m}
\end{align}
Thus, even before analyzing any non-linear dynamics, Eq.~\eqref{eq:minimum_delta_m} yields a model-dependent upper limit on the depth of voids: in theories that never develop a pathology, one can reach $\delta=-1$, while whenever $f_{\rm MG}(z)>1$ voids with $\delta \in[-1,-1/f_{\rm MG}(z)]$ are forbidden at that redshift. 
\begin{figure}[t!]
    \centering
    \includegraphics[width=1.0\linewidth]{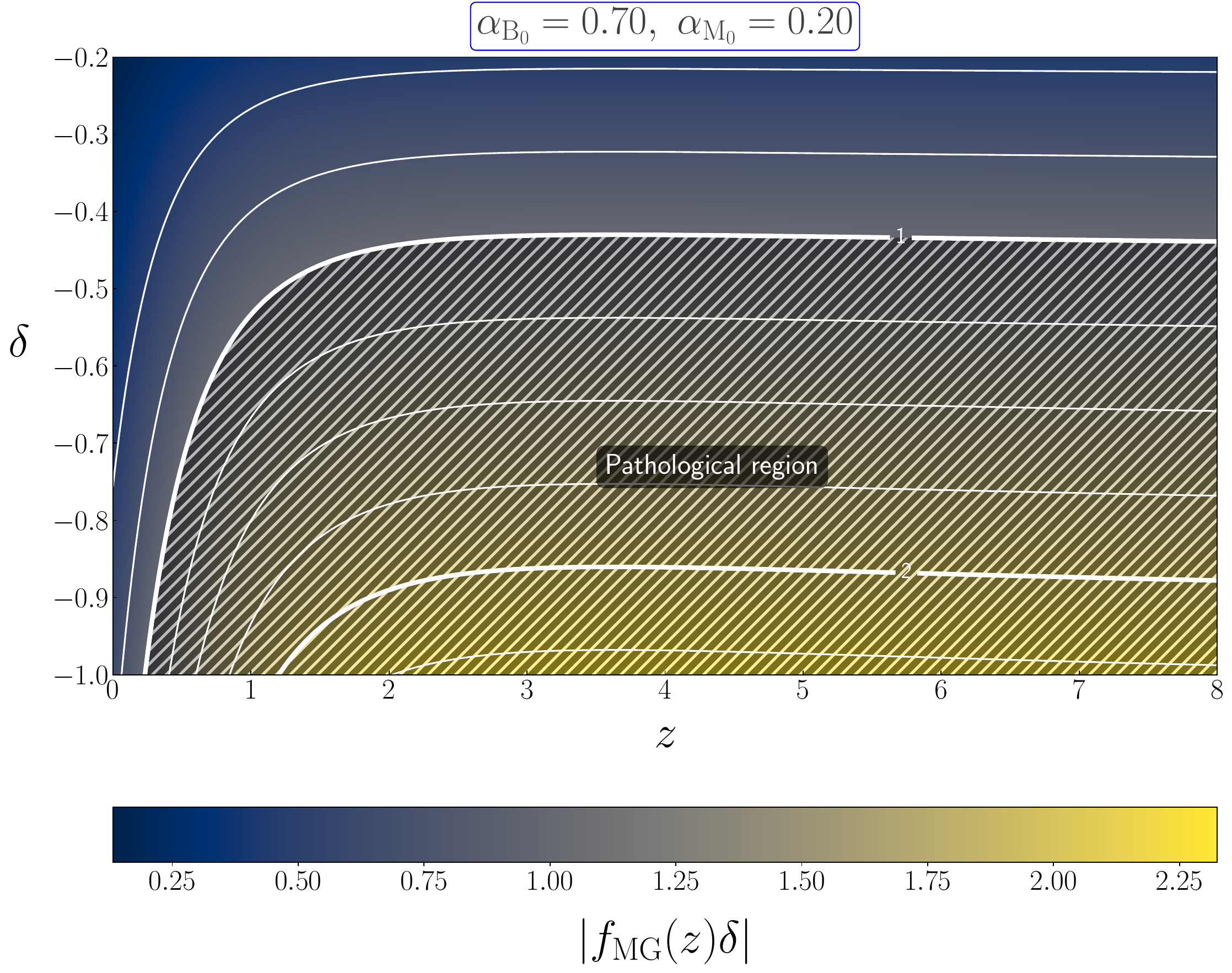}
    \caption{$\lvert f_{\rm MG}(z)\,\delta\rvert$ in the $(z,\delta)$ plane for a Galileon model with $(\alpha_{\rm B_0},\alpha_{\rm M_0})=(0.70,0.20)$. White contours mark selected level sets.}
    \label{fig:single_model_funz_sqrt}
\end{figure}

\section{Results}\label{Sec:results}
We apply the two background-level diagnostics introduced above to assess the viability of Galileon-type models. 
We work on a spatially flat $\Lambda$CDM background, although the same procedure can be extended to DE models with a time-dependent equation of state~\cite{Moretti:2025gbp}. Radiation is neglected because its contribution at the redshifts relevant for structure formation is too small to affect dynamics. The matter component includes both CDM and baryons, described by a single present day density parameter, which we fix to $\Omega_{\rm m,0} = 0.32$. As an illustrative example, we choose a commonly used parametrization linear in the scale factor for $\alpha_{\rm B}$ and $\alpha_{\rm M}$,
\begin{align}
    \alpha_{\rm B}(a) = \alpha_{\rm B_0}\,a\,, \qquad
    \alpha_{\rm M}(a) = \alpha_{\rm M_0}\,a\,,
    \label{eq:parametrization}
\end{align}
where $\alpha_{\rm B_0}$ and $\alpha_{\rm M_0}$ are real constants. We consider values of $(\alpha_{\rm B_0},\alpha_{\rm M_0})$ that are theoretically viable and compatible with current data at the $95\%$ C.L., see Table~2 in~\cite{Noller:2018wyv}. It is natural to expect that the results presented in this work extend to other time parameterizations of $\alpha_{\rm B}$ and $\alpha_{\rm M}$. As an example, in our recent analysis of a different parametrization, the same void-informed viability argument was found to apply as well~\cite{Moretti:2026dfz}.

As a first step, we examine a single representative model to visualize how the pathological regime emerges in the $(z,\delta)$ plane before turning to a systematic exploration of parameter space.
Fig.~\ref{fig:single_model_funz_sqrt} shows the behavior of $\lvert f_{\rm MG}(z)\,\delta\rvert$ for $(\alpha_{\rm B_0},\alpha_{\rm M_0})=(0.70,0.20)$, a choice that lies within the current observational contours and provides an illustrative case study; white contours mark selected level sets, particularly the threshold $\lvert f_{\rm MG}\,\delta\,\rvert=1$ that separates the healthy and pathological regimes. The plot reveals extended regions (hatched areas) where $\lvert f_{\rm MG}\,\delta\,\rvert>1$. The square root argument becomes negative, so $\mu_{\rm NL}$ turns imaginary and the theory is ill defined. The fact that the pathological region extends over density contrasts typical of observed underdense regions shows that the instability is not confined to extremely rare or exotic configurations.

\begin{figure}[t!]
    \centering
    \includegraphics[width=1.0\linewidth]{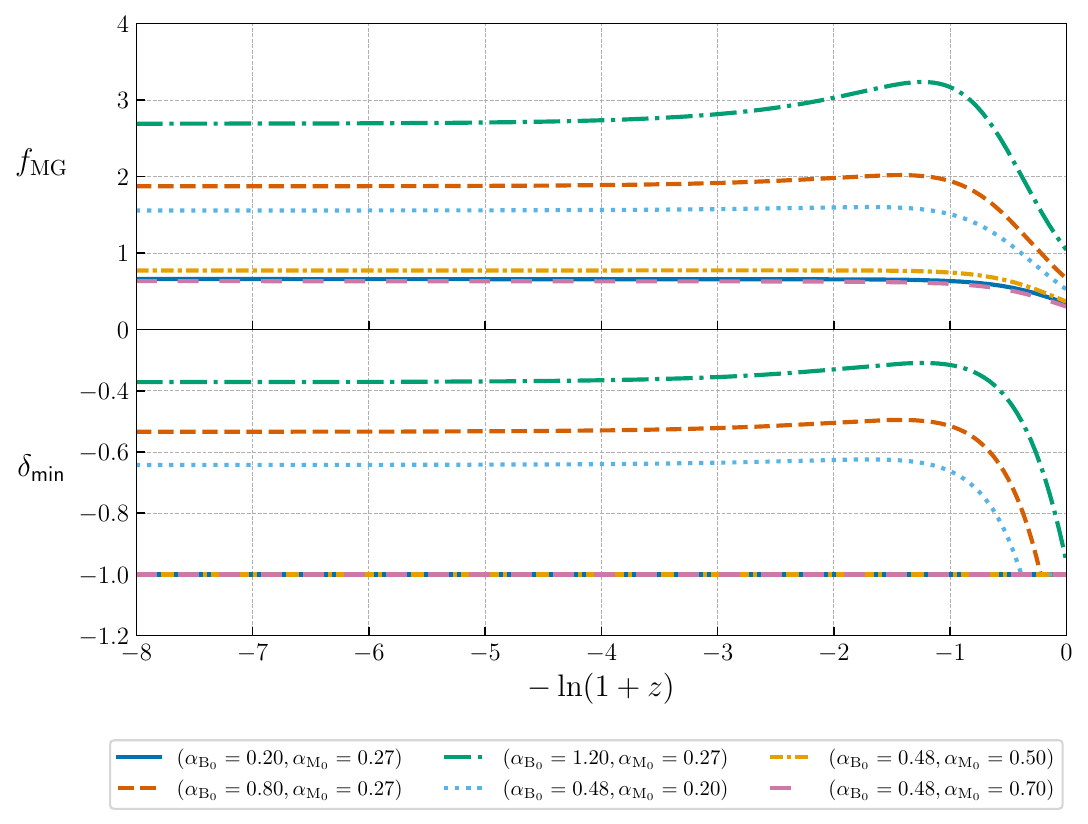}
    \caption{$f_{\rm MG}$ (\textbf{top panel)} and minimum allowed void depth $\delta_{\rm min}(z)$ (\textbf{bottom panel}) as a function of $\ln (1+z)$ for different values of $(\alpha_{\rm B_0},\alpha_{\rm M_0})$.  }
    \label{fig:min_value_delta_m}
\end{figure}

Fig.~\ref{fig:min_value_delta_m} shows $f_{\rm MG}$ (top panel) and $\delta_{\rm min}$ (bottom panel) as a function of $\ln (1+z)\in[0,8]$ for different choices of $(\alpha_{\rm B_0},\alpha_{\rm M_0})$.  We notice that the pathology ($f_{\rm MG}>1$) is widespread across the different parameters considered, and it is not confined to finely tuned corners of parameter space. For several choices of $(\alpha_{\rm B_0},\alpha_{\rm M_0})$, the excluded range in $\delta$ overlaps with density contrasts typical of observed underdense regions ($\delta\in[-0.8,-0.2]$, see, e.g.,~\cite{Verza:2019tvg,Bayer:2021iyb,Verza:2024rbm}), underscoring that this is a phenomenologically relevant issue. For some parameter combinations $\delta_{\rm min}(z)=-1$ throughout the structure-formation era, indicating that the theory does not encounter the imaginary force pathology in the regime of interest ($f_{\rm MG}(z)\leq1$ for any $z\in[0,100]$). For others, $\delta_{\rm min}(z)$ differs significantly from $-1$, showing that a whole interval of deep under-density is forbidden at given epochs. This highlights the power of the method: a single background function $f_{\rm MG}$ is sufficient to compute, with essentially no dynamical input, the minimum under-density allowed by each model as a function of redshift. 
\begin{figure}[t!]
    \centering
    \includegraphics[width=1.0\linewidth]{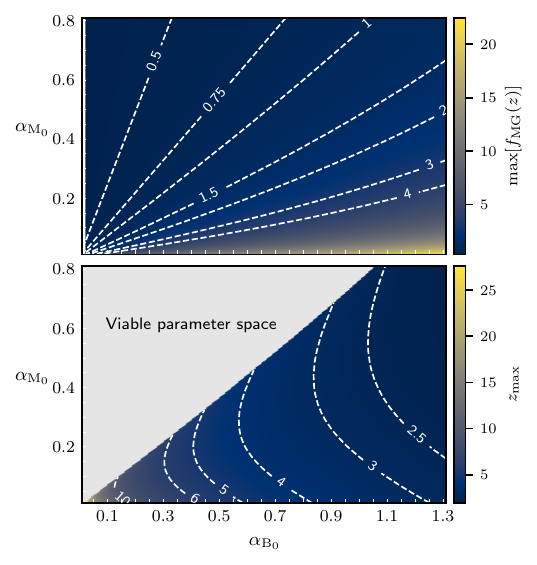}
    \caption{Constraints from Eq.~\eqref{eq:max_criterion} in the $(\alpha_{\rm B_0},\alpha_{\rm M_0})$ plane, spanning the ranges allowed at $95\%$ confidence by~\cite{Noller:2018wyv}. 
    \textbf{Top panel:} Maximum value of $f_{\rm MG}(z)$ over $0\leq z\leq 100$, with dashed contours showing isocontours of $\max f_{\rm MG}$. 
    \textbf{Bottom panel:} Viable region (gray) defined by $\max f_{\rm MG}(z)\leq 1$; outside this area, models are excluded and are colour-coded by the redshift $z_{\rm max}$ at which $f_{\rm MG}$ attains its maximum. }
    \label{fig:max_funz_sqrt}
\end{figure}

In Fig.~\ref{fig:max_funz_sqrt}, we apply the void-informed criterion across the $(\alpha_{\rm B_0},\alpha_{\rm M_0})$ plane, scanning the parameter ranges corresponding to the $95\%$ C.L. reported in~\cite{Noller:2018wyv}, after verifying ghost and gradient stability throughout $z\in[0,100]$. The top panel shows, for each model, the maximum value of $f_{\rm MG}(z)$ attained in the range $0\leq z\leq 100$, with dashed contours indicating lines of constant $\max f_{\rm MG}$. The bottom panel highlights the region of viable parameter space where $\max f_{\rm MG}(z)\leq 1$ is satisfied; outside this gray area, the models are excluded and color-code the redshift $z_{\rm max}$ at which the maximum of $f_{\rm MG}$ occurs. The pathology is thus seen to affect a large fraction of the parameter space, rather than a finely tuned corner: about $60\%$ of the scanned models are ruled out by Eq.~\eqref{eq:max_criterion}, and for almost all of them, the maximum of $f_{\rm MG}$ lies at $z_{\rm max}\lesssim 10$, i.e., squarely within the epoch of structure formation. Thus, using only a background function, one can exclude a substantial portion of the Galileon-type parameter space without ever solving the non-linear dynamics. 

\section{Conclusion}\label{Sec:conclusions}
In this paper, we have introduced a new, void-informed viability condition, which directly applies to Galileon scalar-tensor theories. Because the criterion is expressed solely in terms of background quantities through a single function $f_{\rm MG}$, it can be imposed \textit{a priori}, similarly to ghost/gradient-stability and observational requirements, without solving the non-linear dynamics or performing $N$-body simulations. By enforcing the absence of an imaginary fifth force, this condition provides an efficient additional filter of the Galileon parameter space and excludes models that would otherwise appear viable, while ensuring a physically meaningful description of underdense regions. For the specific parametrization considered in this work, we find that it can be more restrictive than the standard stability bounds in part of the parameter space. Indeed, applying this method to a chosen parametrization of $(\alpha_{\rm B},\alpha_{\rm M})$ within the $95\%$ C.L. region allowed by current constraints, we find that $\sim 60\%$ of the scanned parameter space is ruled out.

We further provide a practical, redshift-dependent prescription to bound the maximum depth of cosmic voids via the minimum allowed density contrast, $\delta_{\rm min}(z)$, thereby quantifying how deep a void can be in a given model. This bound is governed by the same time-dependent background function $f_{\rm MG}$, and determines the maximum void's depth that can be reached without triggering an unphysical (imaginary) fifth force.

Although we have focused on Galileon-type models here, Eqs.~\eqref{eq:max_criterion} and~\eqref{eq:minimum_delta_m} can be extended to any MG theory in which the non-linear force law exhibits the same square-root structure and in which an analogous background function can be identified (as in Eq.~\eqref{eq:non_linear_gravitational_potential}). 

Then, our results highlight cosmic voids as  theory-informed filters for MG, providing a valuable complementary phenomenological motivated priors and a simple diagnostic to guide parameter-space exploration.

\textit{Acknowledgments.}---
We thank the anonymous referee for valuable comments and suggestions that helped improve the manuscript.
T.M. and N.F. acknowledge Baojiu Li for his valuable input. T.M. acknowledges financial support from the Italian Space Agency (ASI) through the ASI-CAIF fellowship.
T.M., N.F. and F.P. acknowledge the Fundação para a Ciência e a Tecnologia (FCT) project with ref. number PTDC/FIS-AST/0054/2021 and the COST Action CosmoVerse, CA21136, supported by COST (European Cooperation in Science and Technology). 
T.M. acknowledges the COST Action BridgeQG, CA23130, supported by COST.
T.M. and N.F. acknowledge the Istituto Nazionale di Fisica Nucleare (INFN) Sez. di Napoli, Iniziativa Specifica InDark.
T.M. and G.V. acknowledge support from the Simons Foundation to the Center for Computational Astrophysics at the Flatiron Institute.
F.P. acknowledges partial support from the INFN grant InDark and from the Italian Ministry of University and Research (\textsc{mur}), PRIN 2022 `EXSKALIBUR - Euclid-Cross-SKA: Likelihood Inference Building for universe's Research', Grant No.\ 20222BBYB9, CUP C53D2300131 0006, and from the European Union -- Next Generation EU.

\bibliographystyle{apsrev4-2}
\bibliography{main}

\end{document}